\documentclass[american, 12pt]{paper}
\usepackage[T1]{fontenc}
\usepackage[utf8]{inputenc}
\usepackage{geometry}
\usepackage{amsthm}
\usepackage{amsmath}
\usepackage{amssymb}
\usepackage[ruled, vlined]{algorithm2e}
\usepackage{crossreftools}
\usepackage{graphicx}
\usepackage[hidelinks]{hyperref}

\makeatletter
\let\paperclassexample\example

\let\example\relax
\AtBeginDocument{%
  \@ifundefined{example}{\let\example\paperclassexample}{}
}

\newtheorem{thm}{\protect\theoremname}
\newtheorem{defn}[thm]{\protect\definitionname}
\newtheorem{prop}[thm]{\protect\propositionname}

\providecommand{\theoremname}{Theorem}
\providecommand{\definitionname}{Definition}
\providecommand{\propositionname}{Proposition}
\providecommand{\lemmaname}{Lemma}
\providecommand{\corollaryname}{Corollary}

\makeatother
\usepackage{babel}
\usepackage{natbib}

\begin{document}
\title{High-dimensional Adaptive MCMC with Reduced Computational Complexity}
\author{
  Max Hird\thanks{\texttt{mhird@uwaterloo.ca}}\\
  Department of Statistics and Actuarial Science\\
  University of Waterloo
  \and
  Samuel Livingstone\\
  Department of Statistical Science\\
  University College London
}
\maketitle
\begin{abstract}
We propose an adaptive MCMC method that learns a linear preconditioner which is dense in its off-diagonal elements but sparse in its parametrisation. Due to this sparsity, we achieve a per-iteration computational complexity of $O(m^2d)$ for a user-determined parameter $m$, compared with the $O(d^2)$ complexity of existing adaptive strategies that can capture correlation information from the target. Diagonal preconditioning has an $O(d)$ per-iteration complexity, but is known to fail in the case that the target distribution is highly correlated, see \citet[Section 3.5]{hird2025a}. Our preconditioner is constructed using eigeninformation from the target covariance which we infer using online principal components analysis on the MCMC chain. It is composed of a diagonal matrix and a product of carefully chosen reflection matrices. On various numerical tests we show that it outperforms diagonal preconditioning in terms of absolute performance, and that it outperforms traditional dense preconditioning and multiple diagonal plus low-rank alternatives in terms of time-normalised performance.
\end{abstract}

\begin{keywords}
  Bayesian Computation, Adaptive MCMC, Online Principal Component Analysis
\end{keywords}

\section{Introduction}\label{sec-intro}

Modern Markov chain Monte Carlo (MCMC) methods must adapt to accommodate the individual features of distributions that interest the modern practitioner.  In Bayesian inference, where MCMC is the work-horse, models are often designed using high dimensional data, which results in high-dimensionality of the Bayesian posterior. In the case of a generalised linear model, for example, the posterior distribution depends on the data $X\in \mathbb{R}^{n \times d}$ through $X\theta$ where $\theta\in\mathbb{R}^d$. If the data is high-dimensional, meaning $d$ is large, then so is the posterior distribution over $\theta$.

Linear preconditioning in MCMC is the act of pushing a distribution through a linear transformation in order to make it easier to sample from. Such a linear transformation is called a `linear preconditioner' or `preconditioner' for short, and is represented using a matrix. Adaptive MCMC is the act of adjusting the Markov chain kernel online to make it more efficient. Many existing adaptive MCMC methods adapt by using information from the chain to learn linear preconditioners (see e.g. \cite{haario2001, tran2024, titsias2023}).  The most commonly sought linear preconditioner is the covariance of the distribution that the MCMC algorithm targets. This is the basis of the adaptive Metropolis algorithm of \citet{haario2001} and is implemented in many popular sampling packages \citep{carpenter2017,abadi2016}. If successful, it has the effect of eliminating inefficiencies in the Markov kernel due to anisotropy in the target that can arise from strong correlations and heterogeneous marginal variances.

In a high-dimensional context, learning and using a preconditioning matrix that is dense in terms of its off-diagonal elements is often infeasible due to the $O\left(d^2\right)$ per-iteration computational complexity. Traditional methods to learn and precondition with the covariance, such as those presented by \citet{haario2001}, are therefore often unusable for high-dimensional targets. One solution is to use only marginal variance information from the target to construct a diagonal preconditioner, see e.g. \citet[Algorithm 5]{andrieu2008}. This results in $O\left(d\right)$ per-iteration computational complexity.  It is known, however, that diagonal preconditioning can perform poorly and in fact be worse than doing nothing in some settings when the target distribution is highly-correlated \citep[Section 3.5]{hird2025a}. 

In this article we introduce a new linear preconditioner that is dense in terms of its off diagonal elements but sparsely parametrised, with a per-iteration computational complexity of $O\left(m^2d\right)$ where $m\in \{1,\ldots,d\}$.  In particular, the method we propose learns the $m$ eigenvectors of the target covariance associated with the largest eigenvalues, along with these eigenvalues. It uses this information to construct a preconditioner that aims to make the target more isotropic by normalising the target scales along these directions. The preconditioner is constructed such that the matrix multiplication, inversion, and square root operations are $O\left(m^2d\right)$. The step in which the preconditioner is learned also incurs $O\left(m^2d\right)$ operations, giving a total iteration complexity of $O\left(m^2 d\right)$.

In Section \ref{subsec:the_ideal_preconditioner} we define our ideal preconditioner (i.e. the preconditioner we propose given full knowledge of the target covariance). We give theoretical results concerning the way in which it isotropises the target. In Section \ref{subsec:construction_of_the_preconditioner_with_reflection_matrices} we lay out the exact parameterisation of the preconditioner given estimates of the top $m$ eigenvalues and eigenvectors of the target covariance.  In Section \ref{subsec:learning_the_preconditioner_with_online_principal_components_analysis} we give the method by which we learn these eigenvalues and eigenvectors. The full implementation of the learning step is given in Section \ref{subsubsec:full_implementation}. Section \ref{sec:experimental_tests} is devoted to the numerical testing of our adaptive scheme, comparing against competing methods such as the `pBaM' method of \citet{modi2025}.

\subsection{Notation}\label{subsec:notation}

We define the set of $d \times d$ positive definite matrices as $\mathrm{PD}_{d\times d}$.  Let $\pi$ be the target distribution on $\mathbb{R}^d$ and define $U:\mathbb{R}^d\to \mathbb{R}$ as its negative log-density. In this paper $U$ is assumed to be twice-differentiable everywhere. We define the target after linear preconditioning with $M\in \mathrm{PD}_{d\times d}$ as the pushforward of $\pi$ through $M^{-1/2}$. We define the condition number of the target after linear preconditioning
with $M$ as
\[
\kappa_{M}:=\frac{\sup_{x\in\mathbb{R}^{d}}\lambda_{1}\left(M^{1/2}\nabla^{2}U\left(x\right)M^{1/2}\right)}{\inf_{x\in\mathbb{R}^{d}}\lambda_{d}\left(M^{1/2}\nabla^{2}U\left(x\right)M^{1/2}\right)}
\]
where $\lambda_{1}\left(A\right)$ and $\lambda_{d}\left(A\right)$
are the leading and trailing eigenvalues of a matrix $A$ respectively.
The condition number of the target is therefore $\kappa_{\mathbf{I}}$.
For $M,N\in\mathrm{PD}_{d\times d}$ we say that $M\preceq N$ whenever
$N-M$ is positive semi-definite. We define the set $O\left(d,m\right):=\left\{ V\in\mathbb{R}^{d\times m}:V^{\top}V=\mathbf{I}_{m}\right\} $.
The normal density with mean $\mu\in\mathbb{R}^{d}$ and covariance
$\Sigma\in\mathrm{PD}_{d\times d}$ at a point $x\in\mathbb{R}^{d}$
is denoted $\mathcal{N}\left(x;\mu,\Sigma\right)$. The Kullback-Leibler (KL)
divergence between two distributions $\mu,\nu\in\mathcal{P}\left(\mathbb{R}^{d}\right)$
is defined as 
\[
KL\left(\mu\left\Vert \nu\right.\right):=\mathbb{E}_{\mu}\left[\log\frac{\mathrm{d}\mu}{\mathrm{d}\nu}\left(X\right)\right]
\]
 whenever $\mu\ll\nu$ and $\infty$ otherwise, where $\mathrm{d}\mu/\mathrm{d}\nu$
is the Radon--Nikodym derivative between $\mu$ and $\nu$. Note that $KL\left(\mu\left\Vert \nu\right.\right) \neq KL\left(\nu\left\Vert \mu\right.\right)$ in general. When the target distribution is in the second argument, we call $KL\left(\mu\left\Vert \pi\right.\right)$ the reverse KL. Finally for $k\in\mathbb{N}$ we define $\left[k\right]$ as the set $\{1,\ldots,k\}$.

\section{A New Linear Preconditioner}

\subsection{The Ideal Preconditioner}\label{subsec:the_ideal_preconditioner}

Define $\{\lambda_i^{\left(\pi\right)}:i\in \left[d\right]\}$ as the set of eigenvalues of the target covariance (in descending order) and $\{v_i^{\left(\pi\right)}:i\in \left[d\right]\}$ the corresponding normalised eigenvectors. Fix $m\in \left[d\right]$. We propose to learn the ideal preconditioner $L = QD \in \mathbb{R}^{d\times d}$ where \[
D = \textup{diag}\left\{\sqrt{\lambda_1^{\left(\pi\right)}}, \sqrt{\lambda_2^{\left(\pi\right)}},\ldots,\sqrt{\lambda_m^{\left(\pi\right)}},1,\ldots,1\right\}
\]and $Q\in O\left(d, d\right)$ has as its first $m$ columns the associated ordered eigenvectors $\{v_i^{\left(\pi\right)}:i\in \left[m\right]\}$. The remaining $d-m$ columns are chosen arbitrarily to make $Q$ orthogonal. We call this preconditioner `ideal' because it can only be constructed in the case that we know the eigeninformation of the target covariance.

\begin{prop}\label{prop:preconditioner_isotropises_target_covariance}
    Assume the target distribution $\pi$ has finite covariance. Let $\tilde{\pi}$ denote the target distribution after preconditioning with $L=QD$ where $D$ and $Q$ are as described above. Then the spectrum of $\textup{Cov}_{\tilde{\pi}}\left(X\right)$ is $\{\underbrace{1, \ldots, 1}_m, \lambda_{m+1}^{(\pi)}, \ldots, \lambda_d^{(\pi)}\}$.
\end{prop}

The proof can be found in Section \ref{proof:preconditioner_isotropises_target_covariance}. Therefore if $\lambda_{m+1}^{\left(\pi\right)}$ and $\lambda_d^{\left(\pi\right)}$ are sufficiently close to one, the target will be made approximately isotropic by the preconditioner $L = QD$. The condition number, as defined in Section \ref{subsec:notation}, measures the anisotropy of the target. A higher condition number indicates a higher level of anisotropy, and a condition number of one indicates that the target is isotropic. It is ubiquitous in the bounds on quantities of interest for MCMC algorithms such as the mixing time and spectral gap, see e.g. \cite[Theorem 49]{andrieu2024}. The bounds on these quantities usually depend on the condition number, where a larger condition number leads to polynomially worse bounds.

\begin{prop}\label{prop:preconditioner_reduces_condition_number}
    Assume $\pi$ has finite covariance $\Sigma_\pi := \textup{Cov}_\pi\left(X\right)$. Let $M \geq m > 0$ be such that $m\mathbf{I}_d \preceq \nabla^2U\left(x\right)\preceq M\mathbf{I}_d$ for all $x \in \mathbb{R}^d$.
    Let $\Sigma = QD^2Q^\top \in \mathrm{PD}_{d\times d}$ where $D\in\mathbb{R}^{d\times d}$ and $Q\in O\left(d, d\right)$ are as described above. Then
    \begin{equation*}
        \frac{\min\left\{\lambda_{d}^{\left(\pi\right)},1\right\}}{\max\left\{\lambda_{m+1}^{\left(\pi\right)},1\right\}}\kappa_{\Sigma_\pi}\leq \kappa_{\Sigma} \leq \frac{\max\left\{\lambda_{m+1}^{\left(\pi\right)},1\right\}}{\min\left\{\lambda_{d}^{\left(\pi\right)},1\right\}}\kappa_{\Sigma_\pi}.
    \end{equation*}
\end{prop}

The proof can be found in Section \ref{proof:preconditioner_reduces_condition_number}. It highlights that if $\lambda_{m+1}^{\left(\pi\right)}$ and $\lambda_d^{\left(\pi\right)}$ are sufficiently close to one, preconditioning with our ideal preconditioner $L = QD$ will have a similar effect to preconditioning with the full covariance matrix of the target.

\subsection{Construction using Reflection Matrices}\label{subsec:construction_of_the_preconditioner_with_reflection_matrices}

Even though we have stated the ideal preconditioner, we must still provide a parametrisation of it that yields a per-iteration complexity of $O\left(d\right)$ in both its learning and use. Computations with the diagonal component $D$ of the preconditioner are straightforward and clearly $O\left(d\right)$. We will focus on the construction of the orthogonal component $Q$ here. Assume that we have a set of $m$ orthonormal vectors $V := \{v_i\in\mathbb{R}^d:i\in \left[m\right]\}$, which can be taken as estimates of the top $m$ eigenvectors of the target covariance.

\begin{defn}
    Given $v,w \in\mathbb{R}^d$ we define the Householder matrix $H\left(v \leftrightarrow w\right)\in O\left(d, d\right)$ with
    \begin{equation*}
        H\left(v \leftrightarrow w\right) := \mathbf{I}_d - 2\frac{\left(v-w\right)\left(v-w\right)^\top}{\left\|v-w\right\|^2_2}.
    \end{equation*}
\end{defn}

Note that $H\left(v \leftrightarrow w\right)$ transforms $v$ into $w$ and vice versa. It also has the properties that $H\left(v \leftrightarrow w\right) = H\left(v \leftrightarrow w\right)^\top$ and $H\left(v \leftrightarrow w\right)^2 = \mathbf{I}_d$. It has determinant $-1$ and is therefore a reflection. We construct the orthogonal component of our preconditioner as $Q := Q_m$ where $Q_m$ is defined iteratively as
\begin{equation}\label{eqn:iterative_Householder_construction}
    \begin{split}
        Q_1 &:= H\left(e_1 \leftrightarrow v_1\right)\\
        Q_k &:= H\left(Q_{k-1}e_k \leftrightarrow v_k\right)Q_{k-1}\textup{ for }k\in\left\{2,\ldots,m\right\}.
    \end{split}
\end{equation}

\begin{prop}\label{prop:householder_columns}
    Fix $m\in \left[d\right]$. If $v_1,\ldots,v_m$ is a set of orthonormal vectors that constructs $Q_m$ as in \eqref{eqn:iterative_Householder_construction}, then the $i$th column of $Q_m$ is $v_i$ for all $i\in \left[m\right]$.
\end{prop}

The proof can be found in Section \ref{proof:householder_columns}.

\subsubsection{Computational Complexity of Matrix Operations}\label{subsubsec:computational_complexity_of_matrix_operations}

The full preconditioner is then $L=QD$ with $Q$ constructed as above, and $D$ a diagonal matrix. Matrix-vector multiplication with either a diagonal or a Householder matrix has $O\left(d\right)$ complexity. Each $Q_k$ contains $Q_{k-1}$, however, for all $k\in \left\{2,\ldots,m\right\}$, meaning to calculate $Q_kv$ we must also calculate $Q_{k-1}v$. For this reason matrix-vector multiplication with $L$ is $O\left(m^2d\right)$.

Calculating the inverse $L^{-1}=D^{-1}Q^\top$ is straightforward as $Q$ is a composition of (symmetric) Householder matrices. In the experiments of Section \ref{sec:experimental_tests} we use the preconditioned Metropolis Adjusted Langevin (MALA) \citep{roberts1996} to compare approaches, meaning such an inverse must be calculated at each accept-reject step to evaluate the proposal density. That $QD$ admits fast inverse computation gives our algorithm a speed advantage over other approaches using a dense preconditioner, despite the fact that we resort to dense matrix-vector multiplication in our experiments rather than exploiting the specific structure of $QD$, see Section \ref{subsubsec:adaptive_schemes} for details. Sampling from $\mathcal{N}\left(0,LL^\top\right)$ can be achieved by multiplying $\xi\sim \mathcal{N}\left(0,\mathbf{I}_d\right)$ by $L$, and so has the same complexity as matrix-vector multiplication.

\section{Learning the Preconditioner with Online Principal Components Analysis}\label{subsec:learning_the_preconditioner_with_online_principal_components_analysis}

\subsection{Eigenvectors}\label{subsubsec:eigenvectors}

Principal components analysis (PCA) is a method for learning the orthogonal directions along which the data exhibit largest variance. Given a dataset $X\in\mathbb{R}^{n \times d}$ whose rows are identically distributed and centred, and a number of directions $m\in\left[d\right]$ we seek
\begin{equation*}
V^* := \textup{argmax}\left\{\mathrm{tr}\left(V^\top \widehat{\Sigma}V\right):V\in \mathbb{R}^{d\times m},V^\top V=\mathbf{I}_m\right\}
\end{equation*}
where $\widehat{\Sigma} := n^{-1}X^\top X$ is used to approximate the covariance. The maximum is achieved when $V^*$ contains the eigenvectors associated with the top $m$ eigenvalues of $\widehat{\Sigma}$. 

PCA can be used on data with any kind of dependence structure between the rows of $X$, but establishing guarantees requires assumptions on the nature of this dependence.  \citet{chen2018} and \citet{kumar2023} provide guarantees for Oja's algorithm \citep{oja1984} in the Markovian setting, and it is therefore our method of choice for Online PCA within our adaptive MCMC learning step. We found that the alternative online PCA method CCIPCA \citep{weng2003} outperforms Oja's algorithm in the $m=1$ case but becomes numerically unstable in the $m>1$ case with Markovian data.

Oja's algorithm is projected gradient descent on the objective $\mathrm{tr}(V^\top\widehat{\Sigma}V)$. The gradient of this objective is $\widehat{\Sigma}V$ and the projection is onto the set $O\left(d,m\right):=\{V\in\mathbb{R}^{d\times m},V^\top V=\mathbf{I}_m\}$. Since the method is online, the full covariance is replaced by the rank one estimate \\$\left(X_t - \mu_t\right)\left(X_t - \mu_t\right)^\top$, giving a gradient estimate of $\left(X_t - \mu_t\right)\left(X_t - \mu_t\right)^\top V$, where $X_t\in\mathbb{R}^d$ is the current state in the Markov chain and $\mu_t\in\mathbb{R}^d$ is a running estimate of the mean of the target distribution. The full update is given by the recursion
\begin{equation*}
    V_t = \Pi_{O\left(d,m\right)}\left(V_{t-1} + \gamma_t\left(X_t - \mu_t\right)\left(X_t - \mu_t\right)^\top V_{t-1}\right)
\end{equation*}
where $\Pi_{O\left(d,m\right)}$ is a projection onto $O\left(d,m\right)$ implemented using a Gram-Schmidt orthogonalisation, and $\gamma_t>0$ is a learning rate.

\citet{cardot2018} recommend using $ct^{-\alpha}$ for the learning rate where $c\in \left(0,\infty\right)$ and $\alpha \in \left(0.5, 1\right]$. After the update columns of $V_t$ serve as the eigenvectors used to construct $Q_m$ as in \eqref{eqn:iterative_Householder_construction}. The product $\left(X_t - \mu_t\right)^\top V_{t-1}$ is $O\left(md\right)$ and the projection with Gram-Schmidt is $O\left(m^2d\right)$.

\subsection{Eigenvalues}\label{subsubsec:eigenvalues}

We now write $Q\left(V\right)$ to denote the orthogonal matrix constructed in
Section \ref{subsec:construction_of_the_preconditioner_with_reflection_matrices}, so that dependence on $V$ is explicit.  If $V$ contains the eigenvectors
corresponding to the top eigenvalues of the target covariance then
\begin{align*}
Q\left(V\right)^{T}\textup{Cov}_\pi\left(X\right)Q\left(V\right) & =\begin{pmatrix}V^{T}\\
W^{T}
\end{pmatrix}\begin{pmatrix}V & W_{\pi}\end{pmatrix}\textup{diag}\left\{ \lambda_{i}^{\pi}:i\in\left[d\right]\right\} \begin{pmatrix}V^{T}\\
W_{\pi}^{T}
\end{pmatrix}\begin{pmatrix}V & W\end{pmatrix}\\
 & =\begin{pmatrix}\mathbf{I}_{m} & 0\\
0 & W^{T}W_{\pi}
\end{pmatrix}\textup{diag}\left\{ \lambda_{i}^{\pi}:i\in\left[d\right]\right\} \begin{pmatrix}\mathbf{I}_{m} & 0\\
0 & W_{\pi}^{T}W
\end{pmatrix}
\end{align*}
where $W\in\mathbb{R}^{d\times\left(d-m\right)}$ is the remaining
$d-m$ columns of $Q\left(V\right)$ and $W_\pi \in\mathbb{R}^{d\times\left(d-m\right)}$ contains the bottom $d - m$ eigenvectors of the target covariance. The second equality holds
because the columns of $V$ are mutually orthonormal, and the columns
of $W$ are orthogonal to the columns of $V$, which are also orthogonal
to the columns of $W_{\pi}$. From the final expression, therefore,
applying $Q\left(V\right)^{T}$ `uncovers' the top $m$ eigenvalues
of the target covariance so that if the MCMC output is scaled
by $Q\left(V\right)^{T}$ the first $m$ marginal variances will be the top $m$ eigenvalues. Defining $\tilde{\mu}_{t}:=Q\left(V_{t}\right)^{T}\mu_{t}$
and $\tilde{X}_{t}:=Q\left(V_{t}\right)^{T}X_{t}$ we propose to learn these eigenvalues using the recursion
\begin{equation*}
    D_{t}=\left(D_{t-1}^{2}+\gamma_{t}\left(\textup{diag}\left\{ \left(\tilde{X}_{t}-\tilde{\mu}_{t}\right)_{i}^{2}:i\in\left[d\right]\right\} -D_{t-1}^{2}\right)\right)^{1/2}
\end{equation*}
where $\gamma_{t} = t^{-\alpha}$ is a learning rate with $\alpha\in\left(0.5, 1\right]$. Then the final $d-m$ diagonal
elements of $D_{t}$ can be set to $1$ in accordance with the description
of the ideal preconditioner in Section \ref{subsec:the_ideal_preconditioner}.

\subsection{Full Implementation}\label{subsubsec:full_implementation}

The full implementation, described in Algorithm \ref{alg:adaptive_step}, has two additional adaptive steps. The first is a standard update
to the adaptive mean $\mu_{t}$ in step 2. The second is an
update of a global scale parameter $\sigma>0$ used in the proposal distribution
of the Markov kernel in step 4.  This is motivated by the optimal scaling literature \citep{beskos2013, roberts1997, rosenthal2011}, which establishes that for suitable high-dimensional targets the optimal value of $\sigma$ can be characterised in terms of the average acceptance probability at equilibrium, independently of the target distribution.

\begin{algorithm}
    \caption{Complete adaptive step}\label{alg:adaptive_step}
    \SetKwInOut{Input}{input}\SetKwInOut{Output}{output}
    \Input{learning rate $\gamma_t > 0$, latest state from the Markov chain $X_t \in \mathbb{R}^d$, latest acceptance probability $\alpha_t\in\left[0,1\right]$, optimal acceptance rate $\alpha^*\in\left(0,1\right]$, latest values of the adaptive parameters $\mu_{t-1}\in \mathbb{R}^d$, $V_{t-1} \in O\left(d,m\right)$, $D_{t-1}\in \mathbb{R}^{d\times d}$, $\sigma_{t-1}>0$, Gram-Schmidt projection operator $\Pi_{O\left(d,m\right)}:\mathbb{R}^{d\times m} \to O\left(d,m\right)$, learning rate coefficient for Oja's algorithm $c \in \left(0,\infty\right)$, general learning rate tuning parameter $\alpha \in \left(0.5, 1\right]$.}
    \Output{updated values of the adaptive parameters $\mu_{t}\in \mathbb{R}^d$, $V_{t} \in O\left(d,m\right)$, $D_{t}\in \mathbb{R}^{d\times d}$, $\sigma_{t}>0$}
    \begin{enumerate}
        \item Set the general learning rate: $\gamma_t = t^{-\alpha}$.
        \item Learn the adaptive mean: $\mu_t = \mu_{t-1} + \gamma_t\left(X_t-\mu_{t-1}\right)$.
        \item Learn the adaptive eigenvector information:
        \begin{enumerate}
            \item Update the eigenvector matrix:
            \begin{equation*}
                \tilde{V}_t=V_{t-1} + c\gamma_t\left(X_t - \mu_t\right)\left(X_t - \mu_t\right)^\top V_{t-1}
            \end{equation*}
            \item Orthogonalise using Gram-Schmidt: $V_t = \Pi_{O\left(d,m\right)}(\tilde{V}_t)$.
        \end{enumerate}
        \item Learn the adaptive global scale:
        \begin{equation*}
            \log \sigma_t = \log \sigma_{t - 1} + \gamma_t\left(\alpha_t - \alpha^*\right).
        \end{equation*}
        \item Learn the adaptive diagonal information
        \begin{enumerate}
            \item Project the location information along the new eigenvectors $V_t$: $\tilde{\mu}_t := Q\left(V_t\right)^\top \mu_t$, $\tilde{X}_t := Q\left(V_t\right)^\top X_t$.
            \item Learn the marginal variances along the new eigenvectors $V_t$:
            \begin{equation*}
                D_{t}=\left(D_{t-1}^{2}+\gamma_{t}\left(\textup{diag}\left\{ \left(\tilde{X}_{t}-\tilde{\mu}_{t}\right)_{i}^{2}:i\in\left[d\right]\right\} -D_{t-1}^{2}\right)\right)^{1/2}
            \end{equation*}
            \item Set $\left(D_t\right)_{ii} = 1$ for $i \in \left\{m+1,\ldots,d\right\}$ in accordance with the ideal preconditioner defined in Section \ref{subsec:the_ideal_preconditioner}.
        \end{enumerate}
    \end{enumerate}
\end{algorithm}

\section{Experiments}\label{sec:experimental_tests}

\subsection{Implementations}\label{subsec:implementations}

Each adaptive scheme uses MALA as its underlying Markov kernel. Pseudocode for this preconditioned kernel can be found in Section \ref{sec:preconditioned_MALA_kernel}. All parameter initialisations and learning rates can be found in the code files in this github repository\footnote{https://github.com/maxhhird/High-dimensional-Adaptive-MCMC-with-Reduced-Computational-Complexity}.

\subsubsection{Details of Adaptive Schemes}\label{subsubsec:adaptive_schemes}

Along with the learning step described in Algorithm \ref{alg:adaptive_step} we also propose a variant in which step 5 (c) is excluded. Excluding step 5 (c) allows the final $d - m$ diagonal elements of $D_t$ to retain information about the marginal scales of the target covariance (along the directions of the final $d - m$ columns of $Q\left(V_t\right)$). We call the variant that includes step 5 (c) `eigen\_identity' and the variant that excludes it `eigen'.  In practice we find simply constructing $Q\left(V_t\right)$, storing it as a matrix, and executing matrix-vector multiplication in the usual way is faster than the bespoke multiplication routines we wrote to exploit its structure. This is because in-built multiplication routines are very well optimised. We therefore use the in-built routines in our numerical experiments. Clearly if the dimension is high enough our bespoke $O\left(m^2d\right)$ multiplications will be faster than the in-built routines (which are $O\left(d^2\right)$).

In all our numerical examples we use $c=1$ and $\alpha = 0.1$ for the $ct^{-\alpha}$ learning rate in the online PCA step of our algorithm. In the $t^{-\alpha}$ learning rates of the other adaptive schemes we choose $\alpha = 0.7$. These choices of hyperparameters yielded the largest step sizes without any instability across all learning updates without the hassle of hand-tuning.

The alternative adaptive schemes that we compare with our `eigen' and `eigen\_identity' strategies are summarised below.

\paragraph{None} The first adaptive algorithm we compare ours to is one where we do no adaptation or preconditioning (apart from adapting the global scale as we note below).

\paragraph{Diagonal} This scheme attempts to learn the marginal standard deviations of the target,
which are then used as the diagonal elements of $L_{t}$. Specifically,
given a new state $X_{t}\in\mathbb{R}^{d}$ and a learning rate $\gamma_{t}>0$
we set
\begin{equation*}
    L_{t}=\left(L_{t-1}^{2}+\gamma_{t}\left(\textup{diag}\left\{ \left(X_{t}-\mu_{t}\right)_{i}^{2}:i\in\left[d\right]\right\} -L_{t-1}^{2}\right)\right)^{1/2}
\end{equation*}
where $\mu_{t}\in\mathbb{R}^{d}$ is a running estimate of the mean
of $\pi$. This learning step is equivalent to that described in Algorithm \ref{alg:adaptive_step} holding $V_{t}$ such that $Q\left(V_{t}\right)=\mathbf{I}_{d}$
and skipping step 5 (c). Matrix multiplication, inversion, and square
root are $O\left(d\right)$ with this preconditioner, and so it provides
the fastest per-iteration algorithm of all those presented here (apart
from `none').

\paragraph{Dense} The final adaptive algorithm for comparison is the dense adaptive scheme
described in \citet[Algorithm 4]{andrieu2008}. The scheme is initialised
at $L_{0}=\mathbf{I}_{d}$ and attempts to learn the target covariance in an online manner. The preconditioner
$L_{t}$ is then set to a matrix square root of the estimate of the
target covariance. Specifically, given a current state $X_{t}\in\mathbb{R}^{d}$
and a learning rate $\gamma_{t}>0$ we set
\begin{equation}\label{eqn:dense_learning_step}
L_{t}=\left(L_{t-1}L_{t-1}^{T}+\gamma_{t}\left(\left(X_{t}-\mu_{t}\right)\left(X_{t}-\mu_{t}\right)^{T}-L_{t-1}L_{t-1}^{T}\right)\right)^{1/2}
\end{equation}
Matrix multiplication and inversion are $O\left(d^{2}\right)$ with
this preconditioner, and so it provides the slowest per-iteration
algorithm of all those presented here. In practice the matrix inside the brackets in \eqref{eqn:dense_learning_step} frequently failed to be positive definite, requiring computation of its minimal eigenvalue in order to restore positive definiteness — an $O(d^3)$ operation.

In each adaptive scheme we also adapt a global scale parameter $\sigma_t$ at each step according to step 4 of Algorithm \ref{alg:adaptive_step}. Each adaptive scheme (including ours) updates all adaptive parameters at every step in the Markov chain.  In all schemes we run two Markov chains simultaneously and, in the learning steps, we average the increments (i.e. the expressions pre-multiplied by the learning rates) across the chains.

\subsubsection{Diagonal Plus Low-rank Preconditioners Learned Pre-chain}

We also compare to alternative preconditioners of the form $D+VV^\top \in \mathbb{R}^{d\times d}$ where $V \in \mathbb{R}^{d\times m}$ and $m \ll d$. These schemes do not use information from the Markov chain to construct their preconditioners; instead the preconditioners are learned before the chain is initialised. 

\paragraph{Reverse Kullback-Leibler Gradient Descent}\label{paragraph:reverse_KL_gradient_descent} The first mechanism we use to learn a diagonal plus low-rank preconditioner is to perform stochastic gradient descent on the objective
\begin{equation*}
KL\left(\mathcal{N}\left(\mu,\left(D+VV^\top\right)\left(D+VV^\top\right)^\top\right)\|\pi\right)
\end{equation*}
where $\mu \in \mathbb{R}^d$, $D\in \mathbb{R}^{d\times d}$ is diagonal and $V \in \mathbb{R}^{d\times m}$. This reverse KL objective is defined in Section \ref{subsec:notation}. Gradients are estimated using Monte Carlo. We call this scheme `diagonal + LR'. Implementation details are given in Section \ref{sec:diagonal_plus_low_rank_gradient_descent_implementation_details}.

\paragraph{Patched Batch and Match (pBaM)}\label{paragraph:patched_batch_and_match} \cite{modi2025} optimise a weighted divergence between the scores of the $\mathcal{N}\left(\mu,\Sigma\right)$ distribution and the target. The descent step uses a batch of $\mathcal{N}\left(\mu,\Sigma\right)$ samples. Each time the covariance parameter is updated it is subsequently projected onto the space of diagonal plus low-rank matrices using an expectation maximisation algorithm. We call this scheme `pBaM'. Each iteration of the `pBaM' algorithm incurs a computational cost that is linear in $d$ but cubic in the batch size.

\subsection{Performance Measure}

We compare samplers using the Effective Sample Size (ESS) (e.g. \cite{plummer2024}). At a high level, the ESS of an MCMC estimator is the number of independent samples needed to achieve a Monte Carlo estimator with the same variance. Note that the ESS depends on expectations with respect to the randomness in the Markov chain at stationarity. We therefore either initialise the chains in stationarity or allow sufficient burn-in to approximate stationarity.  We use the \texttt{effectiveSize} function from the \texttt{coda} package \citep{plummer2024} to estimate the ESS in all cases.

\subsection{Gaussian Targets}\label{subsec:gaussian_targets}

Here we compare `eigen' and `eigen\_identity' (our proposed schemes) with the schemes described in Section \ref{subsec:implementations} on two Gaussian targets with dense, ill-conditioned covariances. The first is tailored to suit our schemes; the second favours methods that learn a diagonal plus low-rank preconditioner.

\subsubsection{Gaussian Target Tailored to Our Method}

The target has the form $\pi=\mathcal{N}(\left(5,\ldots,5\right)^\top,\Sigma_{\pi})$
where $\Sigma_{\pi}$ has $K$ significant eigenvalues sampled from
$\mathcal{N}\left(100,0.01\right)$ and $d-K$ smaller eigenvalues
at $0.1$. This gives the target a condition number of around $1000$.
The first eigenvector of $\Sigma_{\pi}$ is the all ones vector, and
the rest are determined using the \texttt{svd}
function in R. This ensures that $\Sigma_\pi$ is dense in terms of its off-diagonals.

\begin{figure}
    \centering
    \includegraphics[width=0.8\linewidth]{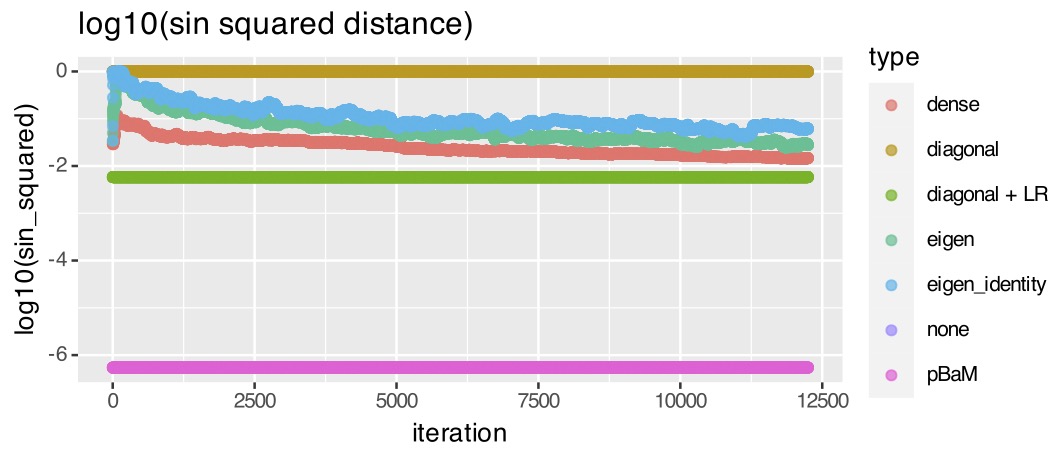}
    \caption{The $\sin^2$ distance (base 10 logarithmic scale) between the preconditioners' leading eigenvector and the leading eigenvector of the target covariance over a single MCMC chain in $d = 150$ with an ill-conditioned Gaussian target.}
    \label{fig:sin_squareds}
\end{figure}

\paragraph{Eigenvector Recovery} First we assess how well each algorithm recovers its optimal parameters. In Figure \ref{fig:sin_squareds}, we compare the $\sin^{2}$
distances between the top eigenvectors of the adaptive preconditioners and the target covariance. We use a $d=150$
dimensional $\mathcal{N}(\left(5,\ldots,5\right)^\top,\Sigma_{\pi})$
target, but with a single significant eigenvalue (meaning $K=1$). %
We examine a single run initialised in equilibrium for $1000\sqrt{d}$
iterations, where the `eigen' and `eigen\_identity'
algorithms learn $m=3$ top eigenvectors, and the rank of the `low-rank' component of the diagonal plus low-rank preconditioners is set to $m = 3$. Adaptive parameter initialisations and learning rates can be found in this repository\footnote{https://github.com/maxhhird/High-dimensional-Adaptive-MCMC-with-Reduced-Computational-Complexity}.

The $\sin^2$ distance for the `none' preconditioner is hidden behind the `diagonal' points; both distances are fixed near 1 for the entire MCMC chain. In the `dense', `eigen', and `eigen\_identity' cases the $\sin^2$ distances drop close to zero over the course of the run. Note that `eigen\_identity' performs slightly worse than the `eigen' algorithm because it ignores the additional marginal variance information, as explained in Section \ref{subsubsec:adaptive_schemes}. The lines for the `diagonal + LR' and `pBaM' are constant because these preconditioners are learned before the chain and fixed during the chain. Both methods accurately recover the top eigenvector of the covariance, particularly the `pBaM' algorithm.

\paragraph{Effective Sample Size} We now increase the number of significant eigenvalues in the target covariance to $K = 3$. The `eigen' and `eigen\_identity'
algorithms learn $m=3$ top eigenvectors, and the rank of the `low-rank' component of the diagonal plus low-rank preconditioners is set to $m = 3$. The algorithms are in dimension $d\in \left\{50,100,150,200\right\}$. The chains are initialised in equilibrium and run for $500\sqrt{d}$ iterations. Adaptive parameter initialisations and learning rates can be found in this repository\footnote{https://github.com/maxhhird/High-dimensional-Adaptive-MCMC-with-Reduced-Computational-Complexity}.

\begin{figure}
    \centering
    \includegraphics[width=0.9\linewidth]{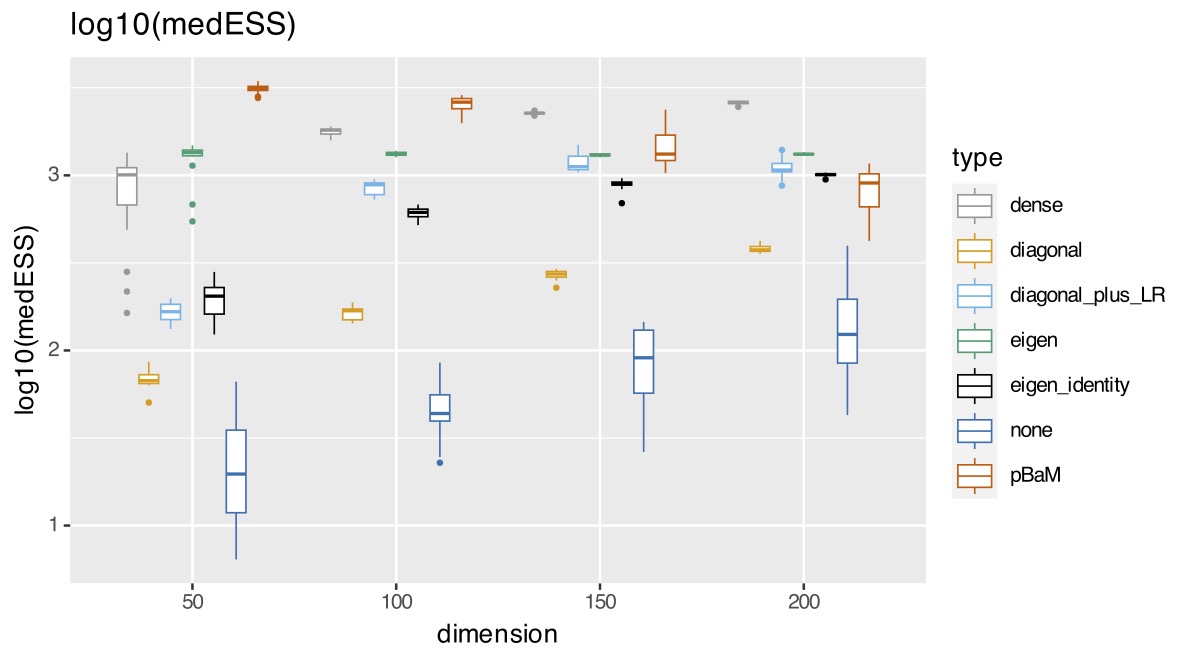}
    \includegraphics[width=0.9\linewidth]{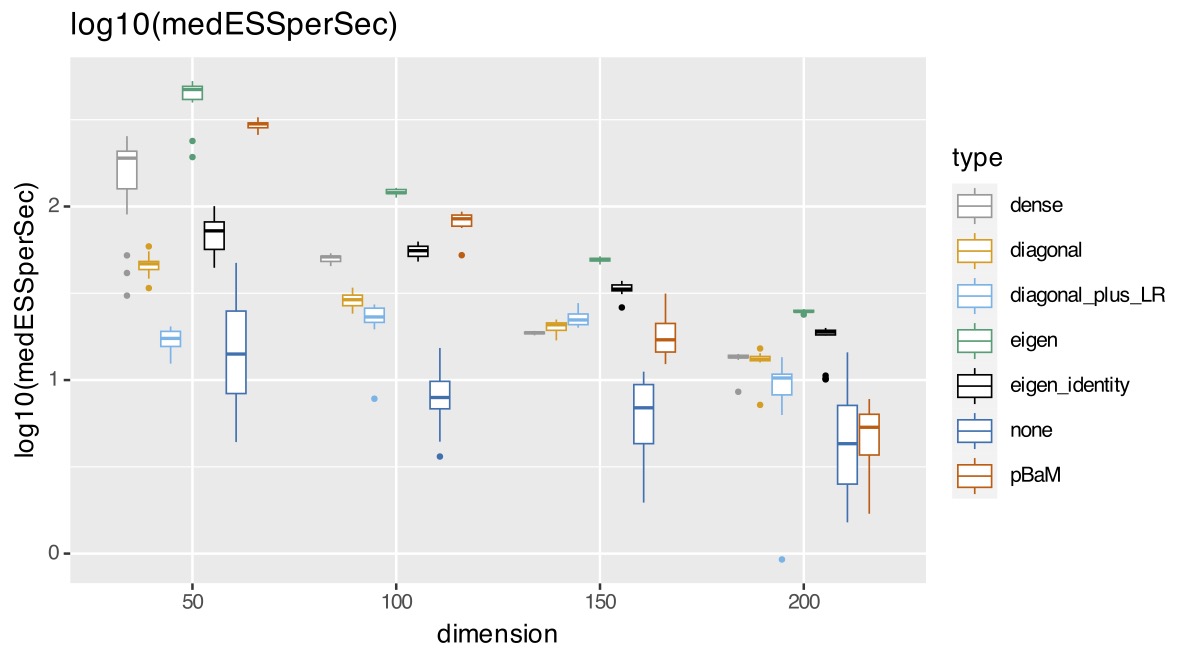}
    \caption{Raw and time-normalised ESSs for adaptive methods on a Gaussian target whose covariance is dense and has $K = 3$ significant eigenvalues in dimension $d\in \left\{50,100,150,200\right\}$.}
    \label{fig:first_Gaussian_ESSs}
\end{figure}

Figure \ref{fig:first_Gaussian_ESSs} shows box plots of the median ESSs over the dimensions of the chain. Each combination of dimension and adaptive scheme is repeated 15 times to achieve the variance displayed in the box plots. The second plot shows the time-normalised ESSs.

From the raw performance (Figure \ref{fig:first_Gaussian_ESSs}, top plot) we see that the `pBaM' method of \citet{modi2025} initially captures the most information about the shape of the target, but then its ability to do so deteriorates in higher dimensions. Secondly we see that the `eigen' method can capture the shape of the target as well as `pBaM' and that its raw performance is constant with the dimension. Thirdly the next most competitive methods are `dense', the `diagonal\_plus\_LR', and the `eigen\_identity', whose performance increases with dimension. The `eigen\_identity' method lags behind the `eigen' method because it ignores the additional marginal variance information, as explained in Section \ref{subsubsec:adaptive_schemes}. The worst schemes are the `diagonal' and `none' schemes because of their failure to capture correlation information.

The per-iteration computational complexity of the `eigen' and `pBaM' schemes scales linearly with dimension and as a consequence these schemes initially have the best time-normalised performance, as shown in the bottom plot of Figure \ref{fig:first_Gaussian_ESSs}.  The raw performance of the `pBaM' scheme decays with dimension, however, and therefore its time-normalised performance decays more rapidly. The quadratic per iteration cost of the `dense' preconditioner means that its dominance in high dimensions in terms of raw ESSs is not carried over to the time-normalised case. In fact, in $d = 200$ it performs similarly to the `diagonal' adaptive method. 

\subsubsection{Gaussian Target Tailored to Diagonal Plus Low Rank Methods}
\label{subsubsec:diaglr}

Here the target is a Gaussian distribution of the form $\mathcal{N}\left(\mu_\pi,D_\pi+V_\pi V_\pi^\top\right)$ where $\mu_\pi \sim \mathcal{N}\left(0,\mathbf{I}_d\right)$, $D_i = \textup{diag}\left\{U_i:i\in \left[d\right]\right\}$ with $U_i \sim U\left[0,1\right]$ independently for all $i\in \left[d\right]$ and $\left(V_\pi\right)_{ij}\sim \mathcal{N}\left(0,1\right)$ for all $i\in \left[d\right]$ and $j \in \left[32\right]$. This replicates the experimental conditions from \citet[Section 5.1.1]{modi2025}. The schemes learning diagonal plus low-rank preconditioners should have an advantage in this case.

\begin{figure}
    \centering
    \includegraphics[width=0.9\linewidth]{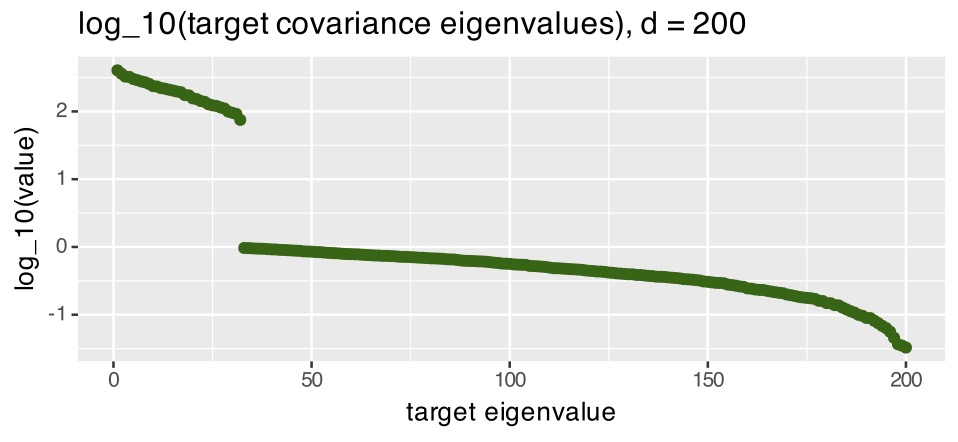}
    \caption{Spectrum of a sample $200 \times 200$ target covariance matrix of Section \ref{subsubsec:diaglr}.}
    \label{fig:diag_plus_LR_evals}
\end{figure}

We are not able to estimate the condition number as precisely as in the first Gaussian experiment, although our evaluations indicate that it increases between approximately ${10}^3$ and ${10}^4$ as dimension increases. Figure \ref{fig:diag_plus_LR_evals} shows the spectrum of a sample target covariance in $200$ dimensions, showing a cluster of $32$ large eigenvalues.

The `eigen'
algorithms learn $m = 10$ top eigenvectors, and the rank of the `low-rank' component of the diagonal plus low-rank preconditioners is set to $32$ with the aim of exactly recovering the target covariance. The algorithms are in dimension $d\in \left\{50,100,150,200\right\}$. The chains are initialised in equilibrium and run for $500\sqrt{d}$ iterations. Adaptive parameter initialisations and learning rates can be found in this github repository\footnote{https://github.com/maxhhird/High-dimensional-Adaptive-MCMC-with-Reduced-Computational-Complexity}. We omit the `eigen\_identity' method because its capability to capture covariance information is dominated by the 'eigen' strategy.

\begin{figure}
    \centering
    \includegraphics[width=0.8\linewidth]{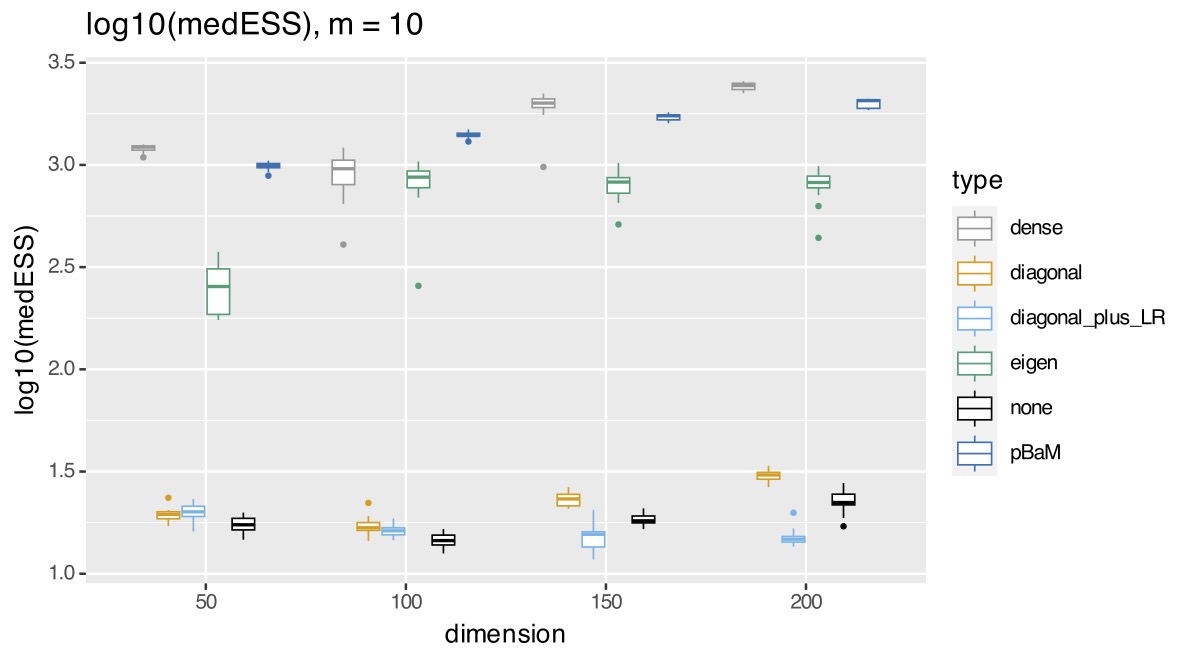}
    \includegraphics[width=0.8\linewidth]{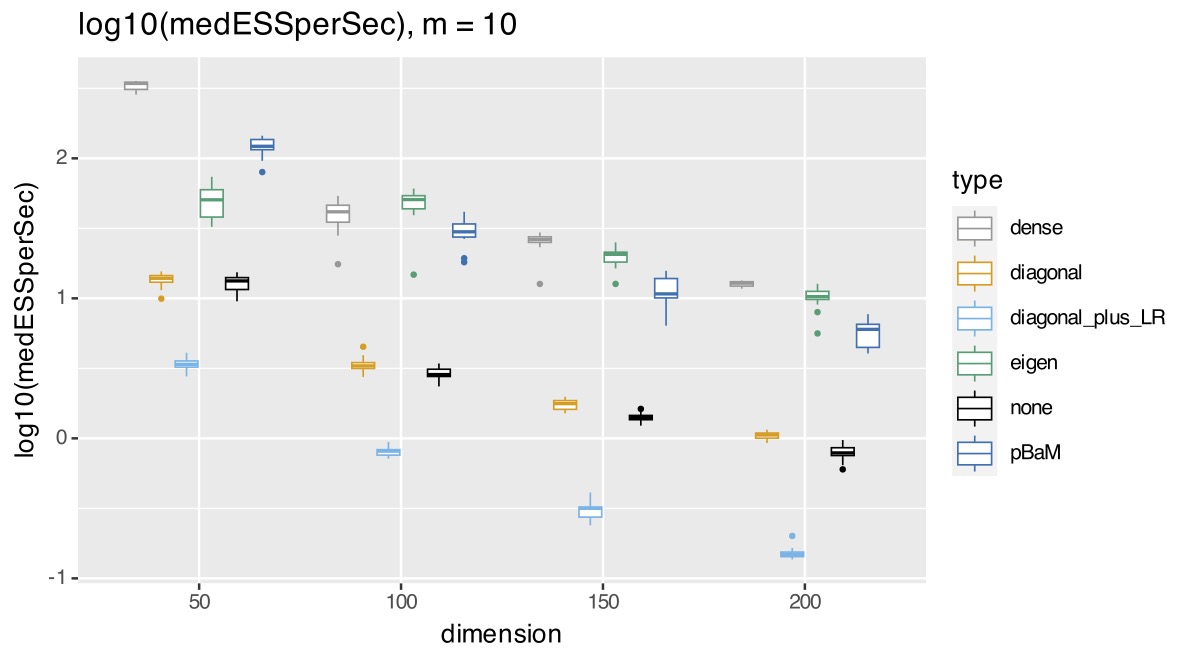}
    \caption{Raw and time-normalised ESSs for adaptive methods on a Gaussian target whose covariance is diagonal plus low-rank in dimension $d\in \left\{50,100,150,200\right\}$.}
    \label{fig:second_Gaussian_ESSs}
\end{figure}

The raw performance is dominated by the `dense' and `pBaM' methods which are trailed by the `eigen' method. Surprisingly, the `diagonal\_plus\_LR' approach does poorly, capturing as much covariance information as the `none' and `diagonal' schemes. Once time-normalised the speed of the `eigen' scheme brings it into competition with the `dense' and `pBaM' approaches, showing that even when the target is tailor made for the diagonal plus low-rank schemes it is still competitive.

\subsection{Bayesian Logistic Regression with Synthetic Data}\label{subsec:bayesian_logistic_regression_with_synthetic_data}

Here we compare the adaptive algorithms on a Bayesian logistic regression posterior with a classical $g$ prior \citep{agliari1988} and synthetic data.

\subsubsection{Experimental Set-up} Here the posterior is proportional to $\exp\left(-U\left(\beta\right)\right)$ where
\begin{equation*}
U\left(\beta\right):=\sum_{i=1}^{n}\left(\left(1-Y_{i}\right)X_{i}^{T}\beta+\log\left(1+\exp\left(-X_{i}^{T}\beta\right)\right)\right)+\frac{\lambda}{2n}\beta^{T}X^{T}X\beta\label{eqn:logistic_regression_posterior}
\end{equation*}
for all $\beta\in\mathbb{R}^d$. Here $Y_{i}$ are the response variables, which are sampled from a \\$\textup{Bernoulli\ensuremath{((1+\exp(-X_{i}^{T}\beta))^{-1})}}$ distribution
for $i\in\left[n\right]$, $X_{i}\in\mathbb{R}^{d}$ are the rows
of the data matrix $X\in\mathbb{R}^{n\times d}$, and $\lambda>0$
is chosen to determine the strength of the prior. The data matrix
is of the form $UDV^{T}$ where $U\in O\left(n, n\right)$ and $V\in O\left(d, d\right)$
are sampled from the Haar measure and $D\in\mathbb{R}^{d\times n}$
is a diagonal matrix with $3$ of the diagonal elements sampled from
a $\mathcal{N}\left(1,10^{-6}\right)$ distribution and the rest set to $\sqrt{1000}$.
The Hessian of the potential is then $\nabla^{2}U\left(\beta\right)=X^{T}\Lambda\left(\beta\right)X$
where
\begin{equation*}
\Lambda\left(\beta\right):=\textup{diag}\left\{ \exp\left(X_{i}^{T}\beta\right)\left(1+\exp\left(X_{i}^{T}\beta\right)\right)^{-2}+\frac{\lambda}{n}:i\in\left[n\right]\right\}.
\end{equation*}
The diagonal elements of $D$ are chosen in an attempt to make $3$
of the eigenvalues of the target covariance much larger than the rest. The condition number
of the posterior is
\begin{equation*}
    \kappa=\kappa\left(X^{T}X\right)\frac{\frac{1}{4}n+\lambda}{\lambda}\approx1000\frac{\frac{1}{4}n+\lambda}{\lambda}.
\end{equation*}
We set $\lambda=0.01$ and $n=d$ for all $d\in\left\{ 50,100,150,200,300\right\} $.
For each algorithmic run we use $k=2$ chains and run for $1000\sqrt{d}$
iterations, measuring the ESSs over the final half of each run. For each combination of dimension and adaptive scheme
we use 15 runs initialised at the mode. Adaptive parameter initialisations and learning rates can be found in this github repository\footnote{https://github.com/maxhhird/High-dimensional-Adaptive-MCMC-with-Reduced-Computational-Complexity}.

\subsubsection{Experimental Results}

\begin{figure}
    \centering
    \includegraphics[width=0.8\linewidth]{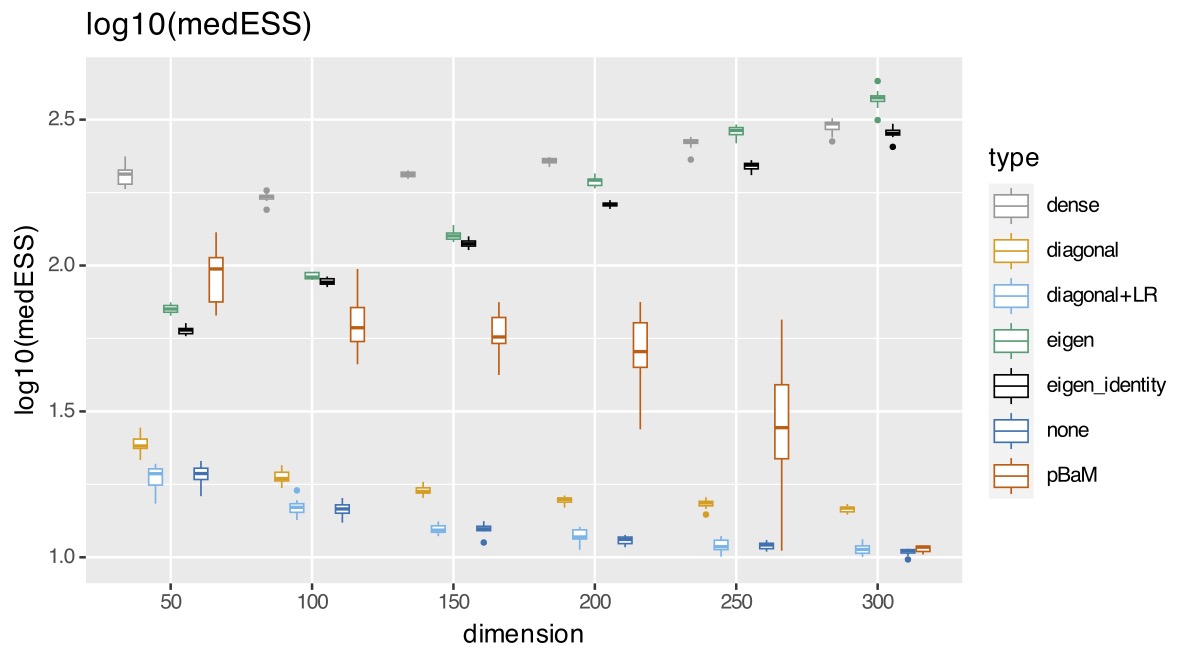}
    \includegraphics[width=0.8\linewidth]{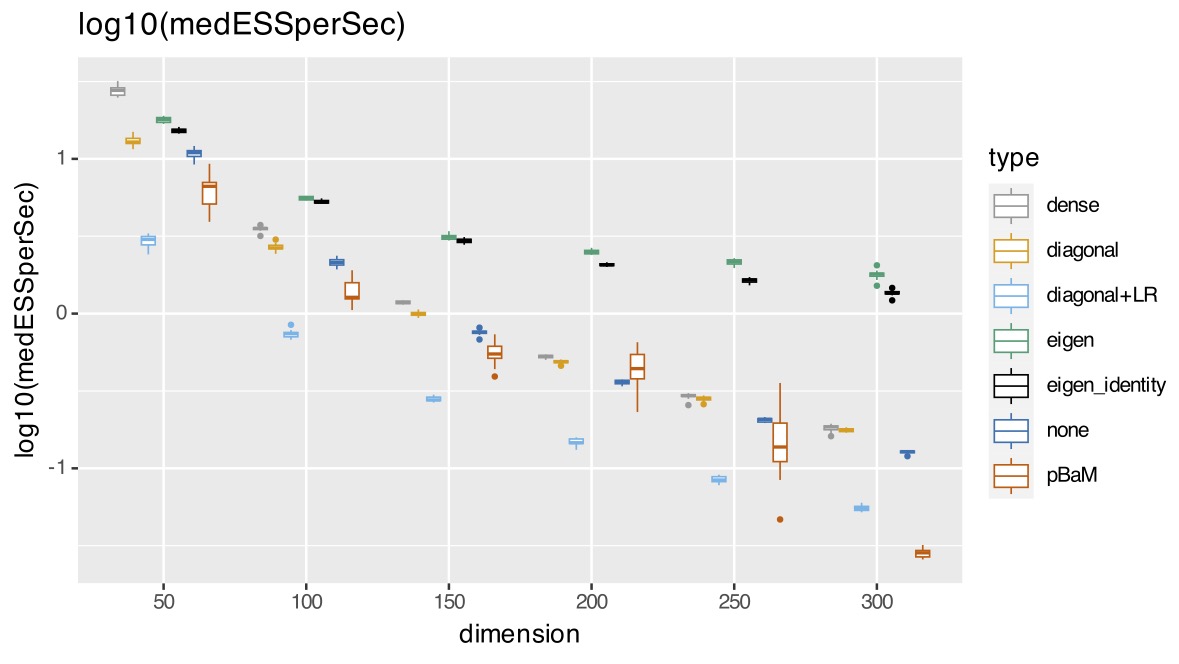}
    \caption{Raw and time-normalised ESSs from the adaptive schemes for a Bayesian logistic regression posterior in various dimensions.}
    \label{fig:logistic_regression_ESSs}
\end{figure}

In Figure \ref{fig:logistic_regression_ESSs} we show the log-transformed medians of the raw ESSs and the log-transformed medians of the time-normalised ESSs across the dimensions of each Markov chain.

As in the Gaussian example in Section \ref{subsec:gaussian_targets} the `eigen\_identity' scheme lags behind the `eigen' scheme in terms of its performance. In this case, however, the differences are not very large. Both schemes remove the effect of the leading eigenvalue of $X^\top X$ on the condition number, but in this case the remaining eigenvalues of $X^\top X$ are 1, so the influence of $\kappa(X^\top X)$ on the condition number cannot be further reduced by the `eigen' scheme, meaning that it loses its advantage over `eigen\_identity'.

The `diagonal' scheme is not significantly better than `none'. This indicates that the target covariance is dense, and has sufficiently many significant off diagonal elements. The raw median ESSs are increasing for both the `dense' and `eigen' schemes, but the computational complexity of the `dense' scheme causes its time-normalised performance to decay sufficiently rapidly that it is dominated by the `eigen' scheme in all dimensions higher than 50.  The raw performance increase is an artifact of the way in which the data matrix $X$ is constructed.

The `diagonal+LR' gradient descent scheme converges but, as shown in Figure \ref{fig:logistic_regression_ESSs}, it does not produce an effective preconditioner. The `pBaM' scheme learning rates are tuned by hand in every case to maximise the speed at which it converges subject to its stability, see the supplementary code files in this github repository\footnote{https://github.com/maxhhird/High-dimensional-Adaptive-MCMC-with-Reduced-Computational-Complexity}.

\subsection{Mean-Field Classical XY Model}\label{subsec:mean_field_classical_XY_model}

The classical XY model describes a system of $d$ planar magnetic spins arranged on a lattice \citep{kirkpatrick2016}.  The state space is $\left[0,2\pi\right]^{d}$, and the potential
energy of a given spin configuration $\theta\in\left[0,2\pi\right]^{d}$
is given by
\begin{equation*}
    U(\theta):=-\frac{1}{2d}\sum_{i,j=1}^{d}J_{ij}\cos(\theta_{i}-\theta_{j})
\end{equation*}
where $J\in\mathbb{R}^{d\times d}$ is a matrix of coupling constants,
and we assume that no external magnetic field acts on the system. We investigate the adaptive schemes on the `mean-field' version of the classical XY model, which has that $J_{ij} = 1$ for all $i,j \in \left[d\right]$.
The distribution over spins is then described by $\pi(d\theta):=Z(\beta)^{-1}\exp(-\beta U(\theta))d\theta$
where $Z(\beta)$ is a normalising constant depending on the inverse
temperature $\beta\in[0,\infty]$.

\begin{figure}
    \centering
    \includegraphics[width=1.0\linewidth]{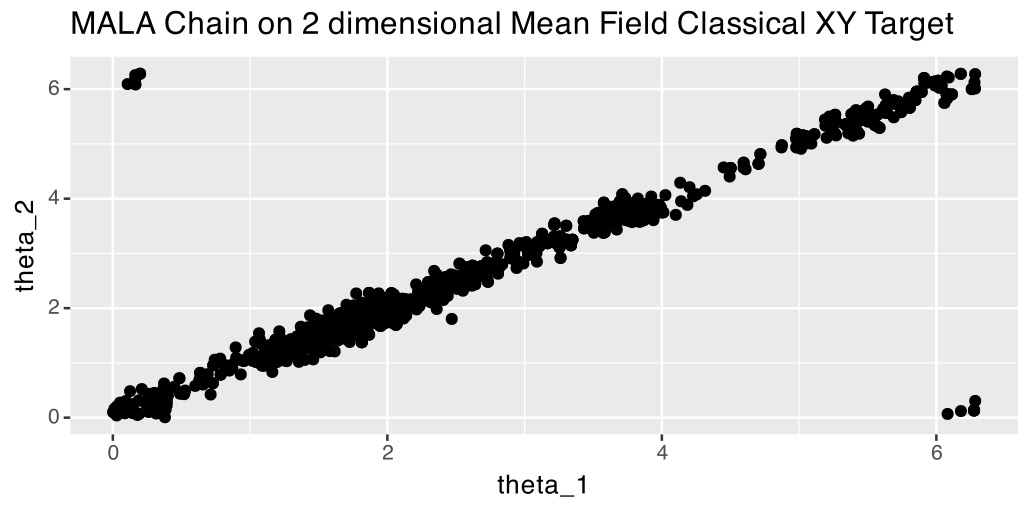}
    \caption{The states of an unpreconditioned MALA chain in $d = 2$ initialised at a mode of the mean-field classical XY model}
    \label{fig:2_dimensional_scatter_plot}
\end{figure}

For a system with $J$ sufficiently
dense in terms of its off-diagonals, we expect many significant correlations
between spin sites. Since $U\left(\theta\right) = U\left(\theta + \lambda \mathbf{1}\right)$ for all $\lambda\in\mathbb{R}$ any minimum $\theta^*$ of $U$ implies infinitely many further minima along the line $\theta^*+ \lambda \mathbf{1}$ and so we expect that the target covariance will have its largest scale along the $\mathbf{1}$ direction. In fact, the set of minima of the potential is exactly $\left\{\lambda \mathbf{1}:\lambda \in \left[0,2\pi\right]\right\}$. Figure \ref{fig:2_dimensional_scatter_plot} shows the states of an unpreconditioned MALA chain in $d = 2$ initialised at a mode of the target at an inverse temperature of $\beta = 100$, corroborating our expectations.

\begin{figure}
    \centering
    \includegraphics[width=0.8\linewidth]{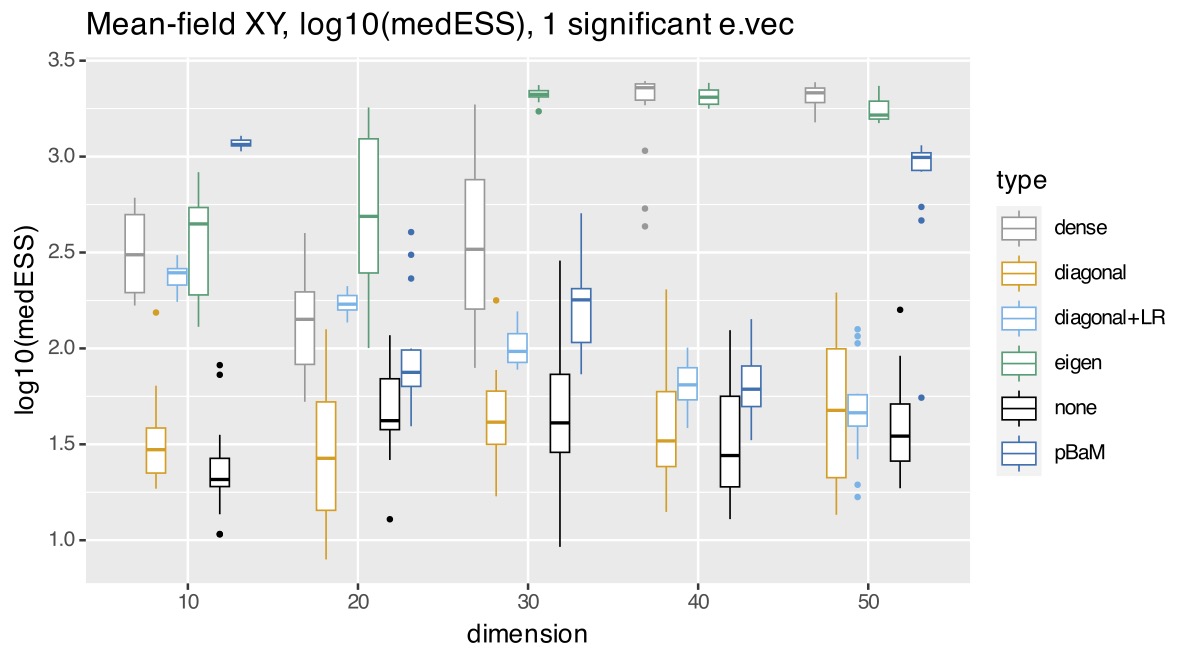}
    \includegraphics[width=0.8\linewidth]{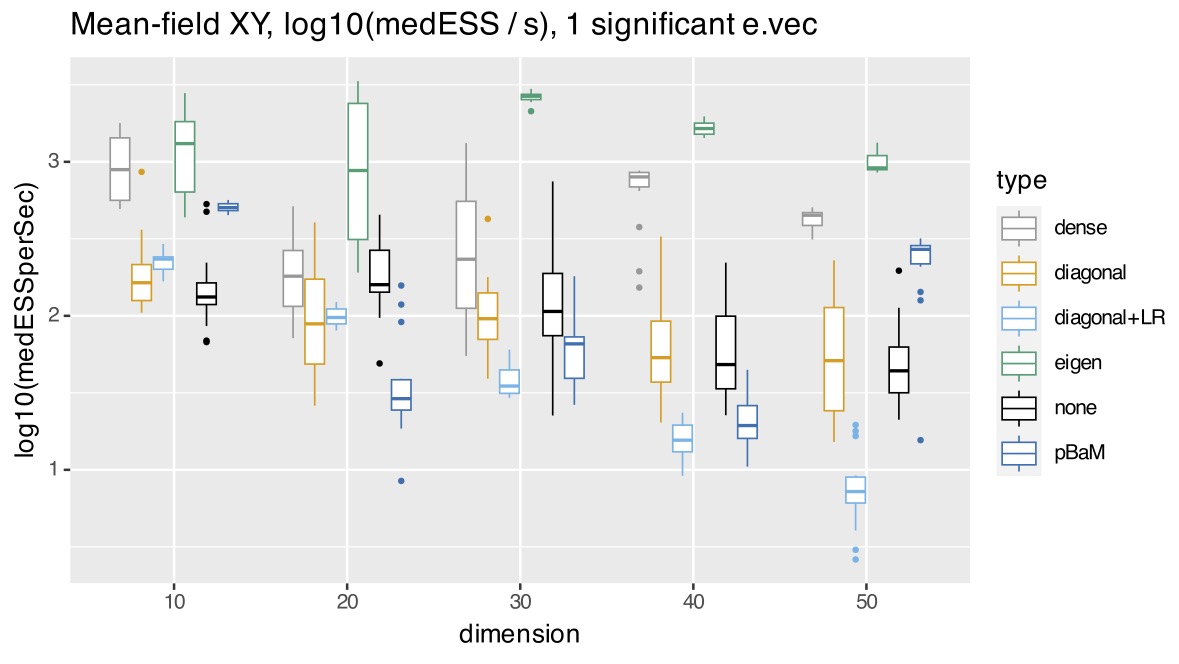}\caption{Raw and time-normalised performances on the mean-field classical XY model for $d \leq 50$. Each boxplot demonstrates the variance in the median of the ESSs over the dimensions of the MCMC run over 15 repetitions.}
    \label{fig:mean_field_classical_XY_ESSs}
\end{figure}

We ran the `dense', `diagonal', `diagonal+LR', `eigen', `none', and `pBaM' schemes on the mean-field classical XY model at an inverse temperature of $\beta = 100$. We initialised at a mode, ran the chains for $1000\sqrt{d}$ iterations, and measured the ESSs over the latter halves of the runs to allow for convergence. Adaptive parameter initialisations and learning rates can be found in this github repository\footnote{https://github.com/maxhhird/High-dimensional-Adaptive-MCMC-with-Reduced-Computational-Complexity}. The results are shown in Figure \ref{fig:mean_field_classical_XY_ESSs}. We neglect to show the results of `eigen\_identity' because it is dominated by the `eigen' scheme. We allowed the `eigen' scheme to learn $m = 1$ eigenvectors and we set the `rank' of the diagonal plus low-rank preconditioners that were learned in the `diagonal+LR' and `pBaM' scheme to 3.  Above $d = 50$, mixing deteriorated to the point where ESS calculations became uninformative.

The set of modes of the target is $\left\{\lambda \mathbf{1}:\lambda \in \left[0, 2\pi\right] \right\}$, which is misaligned with the coordinate axes, meaning in the top plot of Figure \ref{fig:mean_field_classical_XY_ESSs} the `diagonal' and `none' schemes are consistently the worst in terms of their raw ESSs. The preconditioner to which the gradient descent of the `diagonal+LR' scheme converges degrades with dimension, as demonstrated by its deteriorating ESSs. The performance of the `pBaM' scheme varies with dimension, although in each case its computational complexity imposes a large penalty on its time-normalised performance, as can be seen by comparing the relative position of the boxplots between the top and bottom plot of Figure \ref{fig:mean_field_classical_XY_ESSs}. A similar story is true of the `dense' scheme, although the time complexity penalty is more severe in dimensions 40 and 50. The raw performance of the `eigen' scheme is also variable with dimension, but its reduced time complexity makes it the clear frontrunner in dimensions 20 and above. These experiments show that the reduced computational complexity of the `eigen' approach can even convey benefits in low to moderate dimensions.

\section{Conclusion}\label{sec:conclusion}

We introduce a sparsely parametrised preconditioner that uses the top $m$ eigenvalues of the target covariance and their associated eigenvectors, and show that it can be learned adaptively over the course of an MCMC algorithm. It is competitive with standard approaches such as full covariance and diagonal preconditioning, as well as newer methods such as \citet{modi2025} that learn a diagonal plus low-rank preconditioner prior to initialising the chain.  The parametrisation allows key correlations to be captured, while the $O(m^2d)$ per-iteration complexity ensures this comes at a manageable computational cost (subquadratic in dimension when $m < \sqrt{d}$).

\subsection{Extensions and Further Research}

We have left the selection of $m$, the number of eigenvectors to learn, unexamined. One possible selection method is to initialise $m$ at a large value and then gradually reduce it as the eigenstructure of the target covariance is uncovered.  A simplified, $m = 1$ version of our scheme is used within a larger adaptive scheme in \citet[Section 3.2]{riou-durand2023};  replacing it with the full scheme proposed here could improve the overall performance.  We have also restricted attention to the preconditioned MALA kernel. Extending to other base kernels points to the broader question of how the per-sample ESS affects the performance of adaptive MCMC algorithms, particularly those that adapt at every step (e.g. \citet[Section 2]{livingstone2022}). For example, if the per sample ESS of one algorithm is higher than that of another, then this could feed richer information into the adaptive step, potentially creating a virtuous cycle of improving efficiency.

\section{Disclosure statement}\label{disclosure-statement}

The authors declare no conflicts of interest.

\bibliographystyle{plainnat}
\bibliography{bibliography}

\appendix

\section{Proofs}\label{sec:proofs}

\subsection{Proof of Proposition \ref{prop:preconditioner_isotropises_target_covariance}}\label{proof:preconditioner_isotropises_target_covariance}

Let $L=QD\in\mathbb{R}^{d\times d}$ be the ideal preconditioner
and $\Sigma_{\pi}\in\mathbb{R}^{d\times d}$ be the covariance of
$\pi$. This covariance eigendecomposes as $\Sigma_{\pi}=Q_{\pi}D_{\pi}Q_{\pi}^{T}$
where $Q_{\pi}=\left(V_{\pi}\:W_{\pi}\right)\in O\left(d, d\right)$
with $V_{\pi}\in\mathbb{R}^{d\times m}$ having columns $\{ v_{i}^{\left(\pi\right)}:i\in\left[m\right]\} $
and $D_{\pi}=\textup{diag}\{ \lambda_{i}^{\left(\pi\right)}:i\in\left[d\right]\} $.
Since $\Sigma_{\tilde{\pi}}=\left(QD\right)^{-1}\Sigma_{\pi}\left(QD\right)^{-T}=\left(QD\right)^{-1}\Sigma_{\pi}^{1/2}(\left(QD\right)^{-1}\Sigma_{\pi}^{1/2})^{T}$
where $\Sigma_{\pi}^{1/2}:=Q_{\pi}D_{\pi}^{1/2}$ we will work with
$\left(QD\right)^{-1}\Sigma_{\pi}^{1/2}$ because the manipulations
we do will simply be mirrored in $(\left(QD\right)^{-1}\Sigma_{\pi}^{1/2})^{T}$.
We have that
\[
\left(QD\right)^{-1}\Sigma_{\pi}^{1/2}=\begin{pmatrix}\sigma_{\pi}^{-1/2} & 0\\
0 & \mathbf{I}_{d-m}
\end{pmatrix}\begin{pmatrix}V_{\pi}^{T}\\
W^{T}
\end{pmatrix}\begin{pmatrix}V_{\pi} & W_{\pi}\end{pmatrix}D_{\pi}^{1/2}
\]
where $\sigma_{\pi}:=\textup{diag}\{ \lambda_{i}^{\left(\pi\right)}:i\in\left[m\right]\} $
and $W\in\mathbb{R}^{d\times\left(d-m\right)}$ has orthonormal columns,
which are all orthogonal to the columns of $V_{\pi}$. Continuing:
\begin{align*}
\left(QD\right)^{-1}\Sigma_{\pi}^{1/2} & =\begin{pmatrix}\sigma_{\pi}^{-1/2} & 0\\
0 & \mathbf{I}_{d-m}
\end{pmatrix}\begin{pmatrix}V_{\pi}^{T}\\
W^{T}
\end{pmatrix}\begin{pmatrix}V_{\pi} & W_{\pi}\end{pmatrix}D_{\pi}^{1/2}\\
 & =\begin{pmatrix}\sigma_{\pi}^{-1/2} & 0\\
0 & \mathbf{I}_{d-m}
\end{pmatrix}\begin{pmatrix}V_{\pi}^{T}V_{\pi} & V_{\pi}^{T}W_{\pi}\\
W^{T}V_{\pi} & W^{T}W_{\pi}
\end{pmatrix}D_{\pi}^{1/2}\\
 & =\begin{pmatrix}\sigma_{\pi}^{-1/2} & 0\\
0 & \mathbf{I}_{d-m}
\end{pmatrix}\begin{pmatrix}\mathbf{I}_{m} & 0\\
0 & W^{T}W_{\pi}
\end{pmatrix}D_{\pi}^{1/2}\\
 & =\begin{pmatrix}\mathbf{I}_{m} & 0\\
0 & W^{T}W_{\pi}
\end{pmatrix}\begin{pmatrix}\sigma_{\pi}^{-1/2} & 0\\
0 & \mathbf{I}_{d-m}
\end{pmatrix}D_{\pi}^{1/2}\\
 & =\begin{pmatrix}\mathbf{I}_{m} & 0\\
0 & W^{T}W_{\pi}
\end{pmatrix}\begin{pmatrix}\mathbf{I}_{m} & 0\\
0 & \tilde{\sigma}_{\pi}^{1/2}
\end{pmatrix}
\end{align*}
where $\tilde{\sigma}_{\pi}:=\textup{diag}\{ \lambda_{i}^{\left(\pi\right)}:i\in\left\{ m+1,\ldots,d\right\} \} $,
and the third equality comes from the fact that the columns of $V_{\pi}$
are mutually orthogonal, and the columns of $W$ are orthogonal to
the columns $V_{\pi}$ which are also orthogonal to the columns of
$W_{\pi}$. Hence the spectrum of $\Sigma_{\tilde{\pi}}$ is $\{\underbrace{1, \ldots, 1}_m, \lambda_{m+1}^{(\pi)}, \ldots, \lambda_d^{(\pi)}\}$.

\subsection{Proof of Proposition \ref{prop:preconditioner_reduces_condition_number}}\label{proof:preconditioner_reduces_condition_number}

From the proof of Proposition \ref{prop:preconditioner_isotropises_target_covariance} in Section \ref{proof:preconditioner_isotropises_target_covariance} we have that
\[
\Sigma^{-1/2}\Sigma_{\pi}^{1/2}=\begin{pmatrix}\mathbf{I}_{m} & 0\\
0 & W^{T}W_{\pi}
\end{pmatrix}\begin{pmatrix}\mathbf{I}_{m} & 0\\
0 & \tilde{\sigma}_{\pi}^{1/2}
\end{pmatrix}
\]
and hence the spectrum of $\Sigma^{-1/2}\Sigma_{\pi}\Sigma^{-1/2}$
is $\{ 1,\lambda_{m+1}^{\left(\pi\right)},\ldots,\lambda_{d}^{\left(\pi\right)}\} $.
Recall that $U\left(x\right):=-\log\pi\left(x\right)$, and by hypothesis $U$ is twice-differentiable everywhere.  Direct calculation gives
\begin{align*}
\kappa_{\Sigma} & =\frac{\sup_{x\in\mathbb{R}^{d}}\lambda_{1}\left(\Sigma^{1/2}\nabla^{2}U\left(x\right)\Sigma^{1/2}\right)}{\inf_{x\in\mathbb{R}^{d}}\lambda_{d}\left(\Sigma^{1/2}\nabla^{2}U\left(x\right)\Sigma^{1/2}\right)}\\
 & =\frac{\sup_{x\in\mathbb{R}^{d}}\lambda_{1}\left(\Sigma^{1/2}\Sigma_{\pi}^{-1/2}\Sigma_{\pi}^{1/2}\nabla^{2}U\left(x\right)\Sigma_{\pi}^{1/2}\Sigma_{\pi}^{-1/2}\Sigma^{1/2}\right)}{\inf_{x\in\mathbb{R}^{d}}\lambda_{d}\left(\Sigma^{1/2}\Sigma_{\pi}^{-1/2}\Sigma_{\pi}^{1/2}\nabla^{2}U\left(x\right)\Sigma_{\pi}^{1/2}\Sigma_{\pi}^{-1/2}\Sigma^{1/2}\right)}\\
 & \leq\frac{\lambda_{1}\left(\Sigma^{1/2}\Sigma_{\pi}^{-1}\Sigma^{1/2}\right)}{\lambda_{d}\left(\Sigma^{1/2}\Sigma_{\pi}^{-1}\Sigma^{1/2}\right)}\frac{\sup_{x\in\mathbb{R}^{d}}\lambda_{1}\left(\Sigma_{\pi}^{1/2}\nabla^{2}U\left(x\right)\Sigma_{\pi}^{1/2}\right)}{\inf_{x\in\mathbb{R}^{d}}\lambda_{d}\left(\Sigma_{\pi}^{1/2}\nabla^{2}U\left(x\right)\Sigma_{\pi}^{1/2}\right)}\\
 & =\frac{\lambda_{1}\left(\Sigma^{-1/2}\Sigma_{\pi}\Sigma^{-1/2}\right)}{\lambda_{d}\left(\Sigma^{-1/2}\Sigma_{\pi}\Sigma^{-1/2}\right)}\kappa_{\Sigma_{\pi}}\\
 & =\frac{\max\left\{ 1,\lambda_{m+1}^{\left(\pi\right)}\right\} }{\min\left\{ 1,\lambda_{d}^{\left(\pi\right)}\right\} }\kappa_{\Sigma_{\pi}}
\end{align*}
where in the third line we use Ostrowski's Theorem \citep[Theorem 1]{ostrowski1959}.

Similarly we have that 
\begin{align*}
\kappa_{\Sigma_{\pi}} & =\frac{\sup_{x\in\mathbb{R}^{d}}\lambda_{1}\left(\Sigma_{\pi}^{1/2}\nabla^{2}U\left(x\right)\Sigma_{\pi}^{1/2}\right)}{\inf_{x\in\mathbb{R}^{d}}\lambda_{d}\left(\Sigma_{\pi}^{1/2}\nabla^{2}U\left(x\right)\Sigma_{\pi}^{1/2}\right)}\\
 & =\frac{\sup_{x\in\mathbb{R}^{d}}\lambda_{1}\left(\Sigma_{\pi}^{1/2}\Sigma^{-1/2}\Sigma^{1/2}\nabla^{2}U\left(x\right)\Sigma^{1/2}\Sigma^{-1/2}\Sigma_{\pi}^{1/2}\right)}{\inf_{x\in\mathbb{R}^{d}}\lambda_{d}\left(\Sigma_{\pi}^{1/2}\Sigma^{-1/2}\Sigma^{1/2}\nabla^{2}U\left(x\right)\Sigma^{-1/2}\Sigma^{1/2}\Sigma_{\pi}^{1/2}\right)}\\
 & \leq\frac{\lambda_{1}\left(\Sigma_{\pi}^{1/2}\Sigma^{-1}\Sigma_{\pi}^{1/2}\right)}{\lambda_{d}\left(\Sigma_{\pi}^{1/2}\Sigma^{-1}\Sigma_{\pi}^{1/2}\right)}\kappa_{\Sigma}\\
 & =\frac{\lambda_{1}\left(\Sigma_{}^{-1/2}\Sigma_{\pi}\Sigma^{-1/2}\right)}{\lambda_{d}\left(\Sigma_{}^{-1/2}\Sigma_{\pi}\Sigma^{-1/2}\right)}\kappa_{\Sigma}\\
 & =\frac{\max\left\{ 1,\lambda_{m+1}^{\left(\pi\right)}\right\} }{\min\left\{ 1,\lambda_{d}^{\left(\pi\right)}\right\} }\kappa_{\Sigma}
\end{align*}
where in the third line we use Ostrowski's Theorem \citep[Theorem 1]{ostrowski1959}. The result follows
directly.

\subsection{Proof of Proposition \ref{prop:householder_columns}}\label{proof:householder_columns}

We proceed by induction on $m$.

Base Case, $m=1$: Here the set of orthonormal vectors is $\{v_{1}\}$.
It is easily checked that $Q_{1}e_{1}=v_{1}$.

Assume the hypothesis for $m=k$.

Inductive Step, $m=k+1$: Again, it is easily checked that $Q_{k+1}e_{k+1}=v_{k+1}$.
Let $n<k+1$. Then
\begin{align*}
Q_{k+1}e_{n} & =\left(I_{d}-2\frac{(Q_{k}e_{k+1}-v_{k+1})(Q_{k}e_{k+1}-v_{k+1})^{T}}{\|Q_{k}e_{k+1}-v_{k+1}\|^{2}}\right)Q_{k}e_{n}\\
 & =Q_{k}e_{n}-2\frac{(Q_{k}e_{k+1}-v_{k+1})^{T}Q_{k}e_{n}}{\|Q_{k}e_{k+1}-v_{k+1}\|^{2}}(Q_{k}e_{k+1}-v_{k+1})\\
 & =v_{n}-2\frac{e_{k+1}^{T}Q_{k}^{T}Q_{k}e_{n}-v_{k+1}^{T}v_{n}}{\|Q_{k}e_{k+1}-v_{k+1}\|^{2}}(Q_{k}e_{k+1}-v_{k+1})\\
 & =v_{n}
\end{align*}

where $Q_{k}e_{n}=v_{n}$ is true by hypothesis, and $e_{k+1}^{T}Q_{k}^{T}Q_{k}e_{n}=0$
due to the fact that $Q_{k}\in O\left(d, d\right)$.

\section{Preconditioned MALA Kernel}\label{sec:preconditioned_MALA_kernel}

In all our numerical tests the underlying Markov kernel we use for the MCMC algorithms is the preconditioned MALA kernel:

\begin{algorithm} %
    \caption{MALA kernel preconditioned with $L_{t-1}\in\mathbb{R}^{d\times d}$.}
    \SetKwInOut{Input}{input}\SetKwInOut{Output}{output}
    \Input{Previous state $X_{t-1}\in\mathbb{R}^d$, adaptive parameters $L_{t-1}\in\mathbb{R}^{d\times d}$, $\sigma_{t-1}>0$, target density $\pi:\mathbb{R}^d \to \left[0,\infty\right)$ (normalised or unnormalised)}
    \Output{Subsequent state $X_t\in\mathbb{R}^d$}
    \begin{enumerate}
        \item Propose a new state:
        \[
        Y_t=X_{t-1}+\frac{\sigma_{t-1}^2}{2}L_{t-1}L_{t-1}^\top \nabla\log\pi\left(X_{t-1}\right)+\sigma_{t-1}L_{t-1}\xi
        \]
        where $\xi\sim\mathcal{N}\left(0,\mathbf{I}_d\right)$.
        \item Accept the proposed state with probability
        \begin{equation*}
            \begin{split}
                \alpha\left(X_{t-1} \to Y_t\right) &=\\
                \min&\left\{1, \frac{\pi\left(Y_t\right)\mathcal{N}\left(X_{t-1};Y_t + \frac{\sigma_{t-1}^2}{2}L_{t-1}L_{t-1}^\top \nabla\log\pi\left(Y_t\right),\sigma_{t-1}^2L_{t-1}L_{t-1}^\top\right)}{\pi\left(X_{t-1}\right)\mathcal{N}\left(Y_t;X_{t-1}+\frac{\sigma_{t-1}^2}{2}L_{t-1}L_{t-1}^\top \nabla\log\pi\left(X_{t-1}\right),\sigma_{t-1}^2L_{t-1}L_{t-1}^\top\right)}\right\}
            \end{split}
        \end{equation*}
    \end{enumerate}
\end{algorithm}

\section{Diagonal Plus Low Rank Gradient Descent Implementation Details}\label{sec:diagonal_plus_low_rank_gradient_descent_implementation_details}

For the gradient descent scheme we introduce in Section \ref{paragraph:reverse_KL_gradient_descent} we use the following implementation described in Algorithm \ref{alg:reverse_KL_gradient_descent}.

\begin{algorithm}
    \caption{Implementation of the gradient descent scheme on a reverse KL objective between a Gaussian with a diagonal plus low-rank covariance and the target}\label{alg:reverse_KL_gradient_descent}
    \SetKwInOut{Input}{input}\SetKwInOut{Output}{output}
    \Input{Initial diagonal matrix parameter $D_0 = \mathbf{I}_d$, Initial low-rank component $V_0 = 0.1 \times \mathbf{1}_{d \times m}$, batch size $B = 10$, Initial mean parameter $\mu_0 \in\mathbb{R}^d$, number of iterations $N$, mean learning rate $\gamma_\mu>0$, diagonal learning rate $\gamma_\Delta>0$, low-rank component learning rate $\gamma_V > 0$}
    \begin{enumerate}
        \item Let $\Delta \leftarrow \sqrt{D}$.
        \item For $n \in \left[N\right]$ do
        \begin{enumerate}
            \item Calculate $L^{-1} = \left(D + VV^\top\right)^{-1}$ using the Sherman-Morrison-Woodbury formula.
            \item Sample a batch for the Monte Carlo estimators of the gradients \[\left\{X_b\right\}_{b \in \left[B\right]} \sim \mathcal{N}\left(\mu,\left(D+VV^\top\right)\left(D+VV^\top\right)^\top\right)^{\otimes B}.\]
            \item Calculate the $L$ gradient:
            \[
                \nabla_L \leftarrow -L^{-1}-\frac{1}{B}\sum_{b=1}^B \nabla\log\pi\left(X_b\right)X_b^\top.
            \]
            \item Make the descent steps
            \begin{enumerate}
                \item Mean step:
                \[
                \mu \leftarrow \mu - \gamma_\mu \frac{1}{B}\sum_{b=1}^B \nabla\log\pi\left(X_b\right).
                \]
                \item $\Delta$ step: $\Delta \leftarrow \Delta - \gamma_\Delta\Delta \textup{diag}\left(\nabla_L\right)$.
                \item $V$ step: $V \leftarrow V-\gamma_V\left(\nabla_L+\nabla_V^\top \right)V$.
            \end{enumerate}
            \item Let $D \leftarrow \Delta^2$.
        \end{enumerate}
    \end{enumerate}
\end{algorithm}

For particular learning rates, numbers of iterations, and parameter initialisations refer to the supplementary code files in the first author's github\footnote{https://github.com/maxhhird/High-dimensional-Adaptive-MCMC-with-Reduced-Computational-Complexity}.

\end{document}